\documentclass{article}
\usepackage{amsmath}
\usepackage{geometry}

\input{tcilatex}
\begin{document}

\title{Applications of Two Body Dirac Equations to Hadron and Positronium
Spectroscopy}
\author{H. W. Crater\thanks{%
hcrater@utsi.edu}, J. Schiermeyer, J. Whitney \\
The University of Tennessee Space Institute \and C. Y. Wong \\
Oak Ridge National Laboratory}
\maketitle

\begin{abstract}
We review recent applications of the Two Body Dirac equations of constraint
dynamics to meson spectroscopy and describe new extensions to three-body
problems in \ their use in the study of baryon spectroscopy. \ We outline
unique aspects of these equations for QED bound states that distinguish them
among the various other approaches to the relativistic two body problem.\
Finally we discuss recent theorectial solutions of new peculiar bound states
for positronium arising from the Two Body Dirac equations of constraint
dynamics, assuming point particles for the electron and the positron.
\end{abstract}

\section{\qquad Introduction}

The Two-Body Dirac equations (TBDE) of Constraint Dynamics have dual
origins. On the one hand they arise as one of the many quasipotential
reductions of the Beth Salpeter equation (BSE)\cite{saz}. On the other they
arise independently from the development of a consistent covariant approach
to the two-body problem in relativistic classical mechanics independent of
QFT\cite{cra82}. \ In this talk we desribe these two aspects and then go on
to discuss applications to hadron spectroscopy\cite{jim},\cite{yin},\cite%
{whit}. \ The last part of our talk explains the importance we put on
numerical QED tests \cite{becker} of the TBDE and some speculative
theoretical results concerning new positronium states\cite{pec}.

\section{Quasipotential Reduction of the Bethe-Salpeter Equation}

Two body Bethe-Salpeter equation \cite{bse} for spin-zero bound states is%
\footnote{%
The irreducible Bethe-Salpeter kernel $K$ would in general contain charge
renormalization and \ vacuum polarization graphs and could contain
self-energy terms transferred from the inverse propogators.} 
\begin{equation*}
G_{0}^{-1}\Psi =(p_{1}^{2}+m_{1}^{2})(p_{2}^{2}+m_{2}^{2})\Psi =K\Psi .
\end{equation*}%
\ The irreducible kernel $K$ is obtained from the off-mass-shell scattering
amplitude $T$ 
\begin{equation*}
T=K+KG_{0}T,
\end{equation*}%
and would in general contain charge renormalization and \ vacuum
polarization graphs and could contain self-energy terms transferred from the
inverse propogators.

The problems of the two body Bethe-Salpeter equation are its technical
complexity and the existence of abnormal solutions excitations in the
relative time-energy with no proper nonrelativistic limit \cite{wick}, \cite%
{nak}. Recent work with static models has indicated, however, that these
abnormal solutions disappear if one includes all ladder and cross ladder
diagrams \cite{jal}. \ This supports Wick's conjecture on defects of ladder
approximations. \ In the mean time numerous 3D quasipotential reductions of
the Bethe-Salpeter equation had been proposed. In fact, they can be, in
principle, infinite in number\cite{yaes}. \ \ \ \ \ \ \ \ \ \ \ \ \ \ \ \ \
\ \ \ \ \ \ \ \ \ \ \ \ \ \ \ \ \ \ \ \ \ \ \ \ \ \ \ \ \ \ \ \ \ \ \ \ \ \
\ \ \ \ \ \ \ \ \ \ \ \ \ \ \ \ \ \ \ \ \ \ \ \ \ \ \ \ \ \ \ \ \ \ \ \ \ \
\ \ \ \ \ \ \ \ \ \ \ \ \ \ \ \ \ \ \ \ \ \ \ \ \ \ \ \ \ \ \ \ \ \ 

Reductions of the BSE can be obtained from iterating the Bethe Salpeter
equation around a three-dimensional Lorentz invariant hypersurface in
relative momentum ($p$) space. \ This leads to invariant three-dimensional
wave equations for relative motion. The resultant 3D wave equation is not
unique, but depends on the nature of the 3D hypersurface. \ We choose
Todorov's quasipotential equation \cite{tod} which has this Schr\"{o}%
dinger-like form%
\begin{equation}
\left( p^{2}+\Phi \left( x_{1}-x_{2}\right) \right) \psi =b^{2}(w)\psi ,
\label{se}
\end{equation}%
The 3D hypersurface restriction on the relative momentum $p$ $%
=(p_{1}-p_{2})/2$ ($m_{1}=m_{2}$) is defined by 
\begin{eqnarray}
p\cdot P\psi  &=&0, \\
P &=&p_{1}+p_{2}.  \notag
\end{eqnarray}%
This eliminates from the start the problems associated with relative
time/energy. \ \ \ \ \ \ \ \ \ \ \ \ \ \ \ \ \ \ \ \ \ \ \ \ 

\ Defining%
\begin{equation*}
p_{\bot }=p+\hat{P}\cdot p\hat{P},~~p_{\bot }\cdot \hat{P}=0,~\hat{P}=\frac{P%
}{\sqrt{-P^{2}}},~~~~\hat{P}^{2}=-1,
\end{equation*}%
we have%
\begin{equation*}
p^{2}\psi =p_{\bot }^{2}\psi .
\end{equation*}%
The effective eigenvalue in the Schr\"{o}dinger-like equation is 
\begin{equation}
b^{2}=\frac{1}{4w^{2}}[w^{4}-2w^{2}(m_{1}^{2}+m_{2}^{2})+\left(
m_{1}^{2}-m_{2}^{2}\right) ^{2}],  \notag
\end{equation}%
with $w$ the c.m. $\ $invariant energy%
\begin{equation*}
w=\sqrt{-P^{2}}.
\end{equation*}%
The quasipotential $\Phi $ in Eq. (\ref{se}) is related to scattering
amplitude by a Lippmann Schwinger type equation%
\begin{equation}
T-\Phi -\Phi \frac{1}{p_{\bot }^{2}-b^{2}-i0}T=0.  \label{tod}
\end{equation}%
The elastic unitarity condition\cite{tod}, 
\begin{equation*}
T-T^{\dag }=\pi iT\delta (p_{\bot }^{2}-b^{2})T^{\dag },
\end{equation*}%
leads to arbitrariness in the Green function%
\begin{equation*}
\frac{1}{p_{\bot }^{2}-b^{2}-i0}f(p_{\bot }^{2}),~f(b^{2})=1~
\end{equation*}
and to the multiplicity of 3D reductions of BSE. Todorov's choice is $%
f(p_{\bot }^{2})=1$, and his equation displays exact relativistic two-body
kinematics in the absence of interactions, 
\begin{equation*}
w=\sqrt{p^{2}+m_{1}^{2}}+\sqrt{p^{2}+m_{2}^{2}}.
\end{equation*}%
\ The restriction of $p\cdot P\psi =0$ on the time-like component of the
relative momentum is compatible with Eq. (\ref{se}) provided 
\begin{equation*}
\left[ p\cdot P,\Phi \right] \psi =0.
\end{equation*}%
This forces $\Phi $ to depend on $x_{1}-x_{2}$ only through the transverse
component, 
\begin{equation*}
x_{\bot }^{\mu }=\left( \eta ^{\mu \nu }+\hat{P}^{\mu }\hat{P}^{\nu }\right)
(x_{1}-x_{2})_{\nu },~~x_{\bot }\cdot \hat{P}=0.
\end{equation*}%
Thus, in the c.m.\ frame, the hypersurface restriction $p\cdot P\psi =0~$not
only eliminates the relative energy ($p\psi =(0,\mathbf{p)}\psi $) but
implies that the relative time does not appear, i.e. ($x_{\bot }=(0,\mathbf{%
r)}$).

The formal solution of \ Eq. (\ref{tod}) is 
\begin{equation*}
\Phi =T(1+\frac{1}{p_{\bot }^{2}-b^{2}-i0}T)^{-1}.
\end{equation*}%
A nonperturbative approximate solution to this equation has been obtained \ 
\cite{saz97} for this for both world scalar and vector interactions which\
a) includes all ladder and cross ladder diagrams for $T=\sum_{n=1}^{\infty
}T^{(n)}$ and b) includes iterations that result from the geometric series
expansion 
\begin{equation*}
(1+\frac{1}{p_{\bot }^{2}-b^{2}-i0}T)^{-1}=\sum_{m=1}^{\infty }(-)^{m}\left( 
\frac{1}{p_{\bot }^{2}-b^{2}-i0}T\right) ^{m}.
\end{equation*}%
The iterations are called Constraint Diagrams. \ \ For QED-like field
theories, \cite{saz97} uses a scheme that adapts Eikonal approximation for
ladder, cross Ladder, and constraint diagrams to bound states. \ Applied
through all orders it gives for scalar exchange the quasipotential

\begin{equation*}
\Phi =2m_{w}S+S^{2},
\end{equation*}%
while for vector exchange 
\begin{equation*}
\Phi =2\varepsilon _{w}A-A^{2}.
\end{equation*}%
The kinematical variables 
\begin{eqnarray*}
m_{w} &=&\frac{m_{1}m_{2}}{w}, \\
\varepsilon _{w} &=&\frac{w^{2}-m_{1}^{2}-m_{2}^{2}}{2w},
\end{eqnarray*}%
satisfy Einstein relation 
\begin{equation*}
b^{2}=\varepsilon _{w}^{2}-m_{w}^{2},
\end{equation*}%
and corresponds to the energy and reduced mass for the fictitious particle
of relative motion. \ The effects of ladder and cross ladder diagrams thus
embedded in their c.m. energy dependencies

\ By way of the minimal substitutions%
\begin{eqnarray*}
\varepsilon _{w} &\rightarrow &\varepsilon _{w}-A, \\
m_{w} &\rightarrow &m_{w}+S,
\end{eqnarray*}%
one can modify the free two-body equation%
\begin{equation*}
p^{2}\psi =\left( \varepsilon _{w}^{2}-m_{w}^{2}\right) \psi =b^{2}\psi ,
\end{equation*}%
to%
\begin{equation*}
\left( p^{2}+2\varepsilon _{w}A-A^{2}+2m_{w}S+S^{2}\right) \psi =\left(
\varepsilon _{w}^{2}-m_{w}^{2}\right) \psi =b^{2}\psi ,
\end{equation*}%
in which the two particles interact by way of scalar and vector potentials.
\ The form of 
\begin{equation*}
\Phi =2m_{w}S+S^{2}+2\varepsilon _{w}A-A^{2}
\end{equation*}%
is valid for more general potentials than the invariant Coulomb forms 
\begin{equation*}
-\frac{\alpha }{r}=-\frac{\alpha }{\sqrt{x_{\bot }^{2}}},
\end{equation*}%
for which they are derived.

\section{\protect\bigskip Two Body Dirac Equations of Constraint Dynamics}

The Two-Body Dirac equations provide a manifestly covariant 3D reduction of
the BSE for two spin-1/2 particles\cite{cra82}. Furthermore, the constraint
approach \cite{tod78} provides a route around the Currie-Jordan-Sudarshan
\textquotedblleft non-interaction theorem" \cite{cur} which apparently
forbade canonical 4-dimensional treatment of the relativistic $N$- body
problem. \ As with the 3D quasipotential equation, the TBDE covariantly
eliminates relative time and energy. \ For two particles interacting through
scalar and vector interactions the TBDE are given by 
\begin{align*}
\mathcal{S}_{1}\psi & \equiv \gamma _{51}(\gamma _{1}\cdot (p_{1}-\tilde{A}%
_{1})+m_{1}+\tilde{S}_{1})\psi =0, \\
\mathcal{S}_{2}\psi & \equiv \gamma _{52}(\gamma _{2}\cdot (p_{2}-\tilde{A}%
_{2})+m_{2}+\tilde{S}_{2})\psi =0,
\end{align*}%
in which $\psi $ is a 16 component spinor. \ The operators are compatible
with 
\begin{equation*}
\left[ \mathcal{S}_{1},\mathcal{S}_{2}\right] \psi =0,~~\text{implying~}%
\tilde{A}_{i}=\tilde{A}_{i}(x_{\bot }),~\tilde{S}_{i}=\tilde{S}_{i}(x_{\bot
}).
\end{equation*}

One can see the connection to the spin 0 quasipotential results using $%
\varepsilon _{1}$, $\varepsilon _{2}$, the c.m. particle energies 
\begin{eqnarray*}
~\varepsilon _{1}+\varepsilon _{2} &=&w,~\varepsilon _{1}-\varepsilon _{2}=%
\frac{m_{1}^{2}-m_{2}^{2}}{w}, \\
\varepsilon _{1} &=&\frac{1}{2}\left( w+\frac{\left(
m_{1}^{2}-m_{2}^{2}\right) }{w}\right) ,\varepsilon _{2}=\frac{1}{2}\left( w+%
\frac{\left( m_{2}^{2}-m_{1}^{2}\right) }{w}\right) .
\end{eqnarray*}%
Using 
\begin{eqnarray*}
p_{1} &=&\varepsilon _{1}\hat{P}+p,~p_{2}=\varepsilon _{2}\hat{P}-p, \\
p &\equiv &\frac{\varepsilon _{2}p_{1}-\varepsilon _{1}p_{2}}{w},
\end{eqnarray*}%
we rewrite $p\cdot P\psi =0$ and $\left( p^{2}+\Phi \right) \psi
=b^{2}(w)\psi $ as\cite{tod78}%
\begin{eqnarray*}
\mathcal{H}_{1}\psi &=&\left( p_{1}^{2}+m_{1}^{2}+\Phi \right) \psi =0, \\
\mathcal{H}_{2}\psi &=&\left( p_{2}^{2}+m_{2}^{2}+\Phi \right) \psi =0.
\end{eqnarray*}%
The compatibility condition%
\begin{equation*}
\left[ \mathcal{H}_{1},\mathcal{H}_{2}\right] \psi =0,
\end{equation*}%
is satisfied provided that%
\begin{equation*}
\Phi =\Phi (x_{\bot }).
\end{equation*}

For the TBDE, $\ \left[ \mathcal{S}_{1},\mathcal{S}_{2}\right] \psi =0$ also
restricts the spin dependence\ of $\tilde{A}_{i}^{\mu },~\tilde{S}_{i}$ by
determining their dependence on $\gamma _{1},\gamma _{2}$

\ 
\begin{equation*}
\tilde{A}_{i}^{\mu }=\tilde{A}_{i}^{\mu }(A(r),V(r),p_{\perp },\hat{P}%
,w,\gamma _{1},\gamma _{2}),~\ \tilde{S}_{i}=\tilde{S}_{i}(S(r),A(r),p_{%
\perp },\hat{P},w,\gamma _{1},\gamma _{2}).
\end{equation*}%
with vector interactions $\tilde{A}_{i}^{\mu }$ depending on electromagnetic 
$A(r)$ time-like vector $V(r)$ invariant interactions through respective\
vertex forms of $\gamma _{1}\cdot \gamma _{2}$ and $\gamma _{1}\cdot \hat{P}$
$\gamma _{2}\cdot \hat{P}.$ \ Scalar interactions $\ \tilde{S}_{i}$ depend
on scalar invariant $S(r)$ and also vector invariant $A(r).~$\ $~$However, $%
\tilde{S}_{i}(S(r)=0,A(r),p_{\perp },\hat{P},w,\gamma _{1},\gamma _{2})=0.~$%
The\ Pauli reduction of TBDE leads to a covariant Schr\"{o}dinger-like
equation (SLE) for the relative motion with explicit spin-dependent
potential $\Phi .$ \ In the c.m. system: $\ $%
\begin{align}
& \{\mathbf{p}^{2}+\Phi (\mathbf{r,}m_{1},m_{2},w,\mathbf{\sigma }_{1},%
\mathbf{\sigma }_{2})\}\psi _{+}  \notag \\
=& \{\mathbf{p}^{2}+2m_{w}S+S^{2}+2\varepsilon _{w}A-A^{2}+2\varepsilon
_{w}V-V^{2}+\Phi _{D}  \notag \\
& +\mathbf{L\cdot (\sigma }_{1}\mathbf{+\sigma }_{2}\mathbf{)}\Phi _{SO}+%
\mathbf{\sigma }_{1}\mathbf{\cdot \hat{r}\sigma }_{2}\mathbf{\cdot \hat{r}%
L\cdot (\sigma }_{1}\mathbf{+\sigma }_{2}\mathbf{)}\Phi _{SOT}  \notag \\
& +\mathbf{\sigma }_{1}\mathbf{\cdot \sigma }_{2}\Phi _{SS}+(3\mathbf{\sigma 
}_{1}\mathbf{\cdot \hat{r}\sigma }_{2}\mathbf{\ \cdot \hat{r}-\sigma }_{1}%
\mathbf{\cdot \sigma }_{2})\Phi _{T}  \notag \\
& +\mathbf{L\cdot (\sigma }_{1}\mathbf{-\sigma }_{2}\mathbf{)}\Phi _{SOD}+i%
\mathbf{L\cdot \sigma }_{1}\mathbf{\times \sigma }_{2}\Phi _{SOX}\}\psi _{+}
\notag \\
& =b^{2}\psi _{+},  \label{SLE}
\end{align}%
where $\psi _{+}$ is a 4-component spinor subcomponent of 16 component
spinor $\psi $. \ Note that the SLE shares the spin-independent parts
discussed earlier. The TBDE and the equivalent SLE possess important and
desirable features:

\begin{enumerate}
\item TBDE reduce to one-body Dirac form for $m_{1}$ when $m_{2}\rightarrow
\infty $ (the Salpeter equation does not have this property).

\item SLE\ goes into the nonrelativistic Schr\"{o}dinger equation in limit
of weak binding and small speeds.

\item SLE can be solved nonperturbatively for QED bound states of
positronium and muonium as well as QCD meson bound states since: a) every
term in $\Phi $ is less attractive than $-\left( 1/4\right) r^{2}$ (also no $%
\delta (\mathbf{r)}$ or attractive $1/r^{3}$ potentials ) b) the covariant
Dirac formalism introduces natural cutoff factors that smooth out singular
spin-dependent interactions, no need to introduce them by hand as in other
approaches.

\item The $\tilde{A}_{i}^{\mu },~\tilde{S}_{i}$ in the TBDE \ are directly
related to perturbative QFT and for mesons may be introduced
semiphenomenolgically through $A(r)$ and $S(r)$ and $V(r)$.

\item SLE have been tested analytically and numerically against the known
QED perturbative spectrum. The (nonperturbative) successes for the QED
spectrum gives confidence that a numerical treatment of the SLE in QCD
accurately reflects the physical implications of chosen invariant $A,V,S$.

\item TBDE provide covariant 3D framework in which the local potential
approximation consistently fulfills the requirements of gauge invariance in
QED\cite{saz96}.

\item SLE with $\Phi (A=-\alpha /r,V=0,S=0)$ is responsible for accurate QED
spectral results.
\end{enumerate}

For QCD spectra we use $\Phi (A(r)\neq -\alpha /r,V(r)\neq 0,S(r)\neq 0)$
with $A,V,S$ obtained from the static Adler-Piran potential.

\subsection{Two Body Dirac Equations for Meson Spectroscopy}

Adler and Piran \cite{adler} developed a potential for heavy static quarks
from QCD. Their model resembles nonlinear electrostatics with a nonlinear
effective dielectric constant. Integrating their solution fixes all
parameters in their model apart from a mass scale $\Lambda $ and an
``integration constant" $U_{0},$ 
\begin{eqnarray*}
\left( 9/16\pi \right) \nabla \cdot \left[ \ln \left( \mathbf{E}^{2}/\Lambda
^{2}\right) \right] \mathbf{E} &\mathbf{=}&4\pi Q\left[ \delta \left( 
\mathbf{x-x}_{1}\right) -\delta \left( \mathbf{x-x}_{2}\right) \right] , \\
\int \mathbf{E}^{2}(\mathbf{x,x}_{1},\mathbf{x}_{2})d^{3}x
&=&V_{AP}(\left\vert \mathbf{x}_{1}-\mathbf{x}_{2}\right\vert )=\Lambda
(U(\Lambda \left\vert \mathbf{x}_{1}-\mathbf{x}_{2}\right\vert )+U_{0}).
\end{eqnarray*}

\ We divide $V_{AP}$ invariants $A,V$ and $S$ that appear in SLE so that 
\begin{equation*}
V_{AP}(r)+V_{coul}=\Lambda (U(\Lambda r)+U_{0})+\frac{e_{1}e_{2}}{r}=A+V+S\ .
\end{equation*}%
The $V_{AP}$ incorporates asymptotic freedom through 
\begin{equation*}
\Lambda U(\Lambda r<<1)\sim 1/(r\ln \Lambda r),
\end{equation*}%
and confinement through linear and subdominant potential terms, 
\begin{equation*}
\Lambda U(\Lambda r>>1)\sim \Lambda ^{2}r,~~\Lambda \ln \Lambda r,~~\sqrt{%
\frac{\Lambda }{r}},~\frac{k}{r},~a\Lambda .
\end{equation*}

We compute the best fit to the meson spectrum using this division of
Adler-Piran potential\cite{jim}: 
\begin{eqnarray*}
A &=&\exp (-\beta \Lambda r)[V_{AP}-\frac{k}{r}]+\frac{k}{r}+\frac{e_{1}e_{2}%
}{r}, \\
\ \ V+S &=&V_{AP}+\frac{e_{1}e_{2}}{r}-A=(V_{AP}-\frac{k}{r})(1-\exp (-\beta
\Lambda r))\equiv \mathcal{U}.
\end{eqnarray*}%
Thus, $V_{AP}$ is covariantly incorporated into the SLE by treating the
short distance portion as purely electromagnetic-like ($\sim A\gamma _{1\mu
}\gamma _{2}^{\mu }$). The attractive ($k<0$) QCD-Coulomb-like part of $%
V_{AP}\left( \Lambda r>>1\right) $ is assigned completely to
electromagnetic-like\ part $A$. The exponential factor $\exp (-\beta \Lambda
r)$ gradually turns off $A$ at long distances except for $k/r+e_{1}e_{2}./r$
\ The scalar and timelike portions ($S$ and $V$) gradually turn on, becoming
fully responsible for the linear confining and subdominant terms at long
distance. \ The three invariants $A,V,S$ depend on three parameters: $%
\Lambda ,U_{0},$ and $\beta $. \ We introduce a fourth parameter $\xi $
which divides the confining portion $\mathcal{U}$ into scalar and time-like
vector parts: 
\begin{eqnarray*}
e_{2}S &=&\xi \mathcal{U=\xi }(V_{AP}-k/r)(1-\exp (-\beta \Lambda r)), \\
V &=&\mathcal{U-S=(}1-\xi \mathcal{)}(V_{AP}-k/r)(1-\exp (-\beta \Lambda r)).
\end{eqnarray*}

The best fit parameter values are\cite{jim}

$~\ ~~~~~~~~~~~~~~~~~~~~~~~~~~$%
\begin{equation*}
~~~~~~%
\begin{tabular}{|ll|}
\hline
Parameter & Best fit values \\ \hline
$m_{b}$ & $4.953~\text{GeV}$ \\ 
$m_{c}$ & $1.585~\text{GeV}$ \\ 
$m_{s}$ & $0.3079~\text{GeV}$ \\ 
$m_{u}$ & $0.0985~\text{GeV}$ \\ 
$m_{d}$ & $0.1045~\text{GeV}$ \\ 
$\Lambda $ & $0.2255~\text{GeV}$ \\ 
$\Lambda U_{0}$ & $1.770$ GeV \\ 
$\beta \Lambda $ & $0.994~$GeV=$1/(0.198~$fermi) \\ 
$\xi $ & $0.704$ \\ \hline
\end{tabular}%
\end{equation*}%
and indicate that the confining portion begins to dominate at about $0.2$
femis and that the scalar interaction makes up about $70\%$\ $\ $\ of the
confining part of the potential.

The experimental and theoretical values of the meson masses are given in GeV
with the errors given in MeV in parentheses. For $u\bar{d}$ mesons the above
parameters yield

~~~~~~~~%
\begin{equation*}
~~~~~~~~~~~\ 
\begin{tabular}{|lccc|}
\hline
$u\bar{d}$ mesons~~~~~~~~~~~ & Exp. & Th. & $\chi ^{2}$. \\ \hline
$\pi :u\overline{d}\ 1\,{}^{1}S_{0}$ & 0.140(0.0) & 0.134 & 0.3 \\ 
$\rho :u\overline{d}\ 1\,{}^{3}S_{1}$ & 0.775(0.4) & 0.781 & 0.2 \\ 
$b_{1}:u\overline{d}\ 1\,{}^{1}P_{1}$ & 1.230(3.2) & 1.243 & 0.2 \\ 
$a_{1}:u\overline{d}\ 1\,{}^{3}P_{1}$ & 1.230(40.) & 1.320 & 0.1 \\ 
$\pi :u\overline{d}\ 2\,{}^{1}S_{0}$ & 1.300(100) & 1.435 & 0.0 \\ 
$a_{2}:u\overline{d}\ 1\,{}^{3}P_{2}$ & 1.318(0.6) & 1.310 & 0.5 \\ 
$\rho :u\overline{d}\ 2\,{}^{3}S_{1}$ & 1.465(25.) & 1.684 & 0.8 \\ 
$a_{0}:u\overline{d}\ 1\,{}^{3}P_{0}$ & 1.474(19.) & 1.024 & 5.6 \\ 
$b_{2}:u\overline{d}\ 1\,{}^{1}D_{2}$ & 1.672(3.2) & 1.763 & 7.2 \\ \hline
\end{tabular}%
\end{equation*}

Ground state fits are good but some of the radial and some orbital
excitation are off. Note that if we replace 
\begin{tabular}{|lcc}
$a_{0}:u\overline{d}$\ 1\thinspace ${}^{3}P_{0}$ & 1.474(19.) & 1.024 $%
\rangle $%
\end{tabular}
by 
\begin{tabular}{|lcc}
$a_{0}:u\overline{d}$\ 1\thinspace ${}^{3}P_{0}$ & 0.980(20.) & 1.024 $%
\rangle $%
\end{tabular}
the fit is much better. \ In this case we treat the 1.474 GeV meson as a
first radial excitation. \ This leads to 
\begin{tabular}{|lcc}
$a_{0}:u\overline{d}$\ 2\thinspace ${}^{3}P_{0}$ & 1.474(19.) & 1.784 $%
\rangle $.%
\end{tabular}
The fit is better on both accounts.

\ Fits to $s\bar{u},s\bar{d}$ mesons are good for the ground states with
some exceptions on the radial and orbital excitaions. The listed 1.425 meson
would probably be better fit as a radial excitation.

~ 
\begin{equation*}
~~~~~~~~~~~%
\begin{tabular}{|lcccl|}
\hline
$s\bar{u},~s\bar{d}~$Mesons~~~~~~~~~~~~ & Exp. & Th. & $\chi ^{2}$-Th. &  \\ 
\hline
$K\,{}^{-}:s\overline{u}\ 1\,{}^{1}S_{0}$ & 0.494(0.0) & 0.519 & 6.4 &  \\ 
$K\,{}^{0}:s\overline{d}\ 1\,{}^{1}S_{0}$ & 0.498(0.0) & 0.520 & 5.0 &  \\ 
$K^{\ast }\,{}^{-}:s\overline{u}\ 1\,{}^{3}S_{1}$ & 0.892(0.3) & 0.896 & 0.2
&  \\ 
$K^{\ast }\,{}^{0}:s\overline{d}\ 1\,{}^{3}S_{1}$ & 0.896(0.3) & 0.897 & 0.0
&  \\ 
$K\,{}^{-}:s\overline{u}\ 1\,{}^{1}P_{1}$ & 1.272(7.0) & 1.339 & 0.9 &  \\ 
$K^{\ast }\,{}^{-}:s\overline{u}\ 1\,{}^{3}P_{1}$ & 1.403(7.0) & 1.359 & 0.4
&  \\ 
$K^{\ast }\,{}^{-}:s\overline{u}\ 2\,{}^{3}S_{1}$ & 1.414(15.) & 1.706 & 3.8
&  \\ 
$K^{\ast }\,{}^{-}:s\overline{u}\ 1\,{}^{3}P_{0}$ & 1.425(50.) & 1.079 & 0.5
&  \\ 
$K^{\ast }\,{}^{-}:s\overline{u}\ 1\,{}^{3}P_{2}$ & 1.426(1.5) & 1.404 & 1.4
&  \\ 
$K^{\ast }\,{}^{0}:s\overline{d}\ 1\,{}^{3}P_{2}$ & 1.432(1.3) & 1.405 & 2.8
&  \\ 
$K\,{}^{-}:s\overline{u}\ 2\,{}^{1}S_{0}$ & 1.460(40.) & 1.476 & 0.0 &  \\ 
$K^{\ast }\,{}^{-}:s\overline{u}\ 1\,{}^{3}D_{1}$ & 1.717(27.) & 1.837 & 0.2
&  \\ 
$K\,{}^{-}:s\overline{u}\ 1\,{}^{1}D_{2}$ & 1.773(8.0) & 1.803 & 0.1 &  \\ 
$K^{\ast }\,{}^{-}:s\overline{u}\ 1\,{}^{3}D_{3}$ & 1.776(7.0) & 1.792 & 0.0
&  \\ 
$K^{\ast }\,{}^{-}:s\overline{u}\ 1\,{}^{3}D_{2}$ & 1.816(13.) & 1.795 & 0.0
&  \\ \hline
\end{tabular}%
\end{equation*}

\bigskip

The $s\bar{s}$ family of mesons shows a good fit to the ground state and
usual mix of results to the spin-orbit triplet. 
\begin{equation*}
~~~~\ 
\begin{tabular}{|lccc|}
\hline
$s\bar{s}$ Mesons~~~~~~~~~~~~ & Exp. & Th. & $\chi ^{2}$-Th. \\ \hline
$\phi :s\overline{s}\ 1\,{}^{3}S_{1}$ & 1.019(0.0) & 1.013 & 0.4 \\ 
$\phi :s\overline{s}\ 1\,{}^{3}P_{0}$ & 1.370(100) & 1.175 & 0.0 \\ 
$\phi :s\overline{s}\ 1\,{}^{3}P_{1}$ & 1.518(5.0) & 1.437 & 2.5 \\ 
$\phi :s\overline{s}\ 1\,{}^{3}P_{2}$ & 1.525(5.0) & 1.506 & 0.1 \\ 
$\phi :s\overline{s}\ 2\,{}^{3}S_{1}$ & 1.680(20.) & 1.875 & 0.9 \\ 
$\phi :s\overline{s}\ 1\,{}^{3}D_{3}$ & 1.854(7.0) & 1.879 & 0.1 \\ 
$\phi :s\overline{s}\ 2\,{}^{3}P_{2}$ & 2.011(70) & 2.128 & 0.0 \\ \hline
\end{tabular}%
\end{equation*}

The $c\bar{u},c\bar{d}.$ and $c\bar{s}$ mesons display good fits for ground
states

\begin{equation*}
\begin{tabular}{|lccc|}
\hline
$c\bar{u},~c\bar{d},~c\bar{s}~\ $Mesons~~~~~~~~~~~ & Exp. & Th. & $\chi ^{2}$%
-Th. \\ \hline
$D^{0}:c\overline{u}\ 1\,{}^{1}S_{0}$ & 1.865(0.2) & 1.876 & 1.1 \\ 
$D^{+}:c\overline{d}\ 1\,{}^{1}S_{0}$ & 1.870(0.2) & 1.883 & 1.7 \\ 
$D^{\ast 0}:c\overline{u}\ 1\,{}^{3}S_{1}$ & 2.007(0.2) & 2.007 & 0.0 \\ 
$D^{\ast +}:c\overline{d}\ 1\,{}^{3}S_{1}$ & 2.010(0.2) & 2.013 & 0.1 \\ 
$D^{\ast 0}:c\overline{u}\ 1\,{}^{3}P_{0}$ & 2.352(50.) & 2.221 & 0.1 \\ 
$D^{\ast +}:c\overline{d}\ 1\,{}^{3}P_{0}$ & 2.403(14.) & 2.230 & 1.5 \\ 
$D^{+}:c\overline{d}\ 1\,{}^{3}P_{2}$ & 2.460(3.0) & 2.414 & 2.1 \\ 
$D^{\ast 0}:c\overline{u}\ 1\,{}^{3}P_{2}$ & 2.461(1.6) & 2.409 & 7.7 \\ 
\hline
\end{tabular}%
\end{equation*}%
Note that the $c\bar{s}$ spin orbit triplet gives reasonable fits as opposed
to $s\bar{s}$\ 

\begin{equation*}
\begin{tabular}{|lccc|}
\hline
$D_{s}:c\overline{s}\ 1\,{}^{1}S_{0}$ \ \ \ \ \ \ \ \ \ \ \ \ \ \ \ \ \ \ \
\  & 1.968(0.3) & 1.974 & 0.3 \\ 
$D_{s}^{\ast }:c\overline{s}\ 1\,{}^{3}S_{1}$ & 2.112(0.5) & 2.119 & 0.4 \\ 
$D_{s}^{\ast }:c\overline{s}\ 1\,{}^{3}P_{0}$ & 2.318(0.6) & 2.340 & 3.5 \\ 
$D_{s}:c\overline{s}\ 1\,{}^{1}P_{1}$ & 2.535(0.3) & 2.499 & 11.6 \\ 
$D_{s}^{\ast }:c\overline{s}\ 1\,{}^{3}P_{2}$ & 2.573(0.9) & 2.532 & 8.9 \\ 
$D_{s}^{\ast }:c\overline{s}\ 2\,{}^{3}S_{1}$ & 2.690(7.0) & 2.702 & 0.0 \\ 
\hline
\end{tabular}%
\end{equation*}%
\ 

The charmonium family is given by 
\begin{equation*}
\ 
\begin{tabular}{|lccc|}
\hline
$c\bar{c}$ Mesons~~~~~~~~~~~~ & Exp. & Th. & $\chi ^{2}$-Th. \\ \hline
$\eta _{c}:c\overline{c}\ 1\,{}^{1}S_{0}$ & 2.980(1.2) & 2.973 & 0.2 \\ 
$J/\psi (1S):c\overline{c}\ 1\,{}^{3}S_{1}$ & 3.097(0.0) & 3.128 & 9.7 \\ 
$\chi _{0}:c\overline{c}\ 1\,{}^{3}P_{0}$ & 3.415(0.3) & 3.397 & 3.0 \\ 
$\chi _{1}:c\overline{c}\ 1\,{}^{3}P_{1}$ & 3.511(0.1) & 3.505 & 0.4 \\ 
$h_{1}:c\overline{c}\ 1\,{}^{1}P_{1}$ & 3.526(0.3) & 3.523 & 0.1 \\ 
$\chi _{2}:c\overline{c}\ 1\,{}^{3}P_{2}$ & 3.556(0.1)) & 3.557 & 0.0 \\ 
$\eta _{c}:c\overline{c}\ 2\,{}^{1}S_{0}$ & 3.637(4.0) & 3.602 & 0.7 \\ 
$\psi (2S):c\overline{c}\ 2\,{}^{3}S_{1}$ & 3.686(0.0) & 3.689 & 0.1 \\ 
$\psi (1D):c\overline{c}\ 1\,{}^{3}D_{1}$ & 3.773(0.4) & 3.807 & 0.9 \\ 
$\chi _{2}:c\overline{c}\ 2\,{}^{3}P_{2}$ & 3.929(5.0) & 3.983 & 1.1 \\ 
$\psi (3S):c\overline{c}\ 3\,{}^{3}S_{1}$ & 4.039(10.) & 4.092 & 0.3 \\ 
$\psi (2D):c\overline{c}\ 2\,{}^{3}D_{1}$ & 4.153(3.0) & 4.169 & 0.3 \\ 
$\psi (4S):c\overline{c}\ 4\,{}^{3}S_{1}$ & 4.421(4.0) & 4.426 & 0.0 \\ 
$\psi (3D):c\overline{c}\ 3\,{}^{3}D_{1}$ & 4.421(4.0) & 4.483 & 2.3 \\ 
\hline
\end{tabular}%
\end{equation*}%
The overall fit is good with the worst fit meson of the family is the $%
J/\psi .$ \ The $b\bar{u},~b\bar{d},$ $b\bar{s},b\bar{c}~$mesons 
\begin{equation*}
\begin{tabular}{|lccc|}
\hline
$b\bar{u},~b\bar{d}$ $b\bar{s}~$Mesons~~~~~~~~~~~~ & Exp. & Th. & $\chi ^{2}$%
-Th. \\ \hline
$B^{-}:b\overline{u}\ 1\,{}^{1}S_{0}$ & 5.279(0.3) & 5.283 & 0.2 \\ 
$B^{0}:b\overline{d}\ 1\,{}^{1}S_{0}$ & 5.280(0.3) & 5.284 & 0.2 \\ 
$B^{\ast -}:b\overline{u}\ 1\,{}^{3}S_{1}$ & 5.325(0.5) & 5.333 & 0.5 \\ 
$B^{\ast -}:b\overline{u}\ 1\,{}^{3}P_{2}$ & 5.747(2.9) & 5.687 & 3.8 \\ 
$B_{s}^{0}:b\overline{s}\ 1\,{}^{1}S_{0}$ & 5.366(0.6) & 5.367 & 0.0 \\ 
$B_{s}^{\ast 0}:b\overline{s}\ 1\,{}^{3}S_{1}$ & 5.413(1.3) & 5.430 & 1.0 \\ 
$B_{s}^{\ast 0}:b\overline{s}\ 1\,{}^{3}P_{1}$ & 5.829(0.7) & 5.792 & 9.4 \\ 
$B_{s}^{\ast 0}:b\overline{s}\ 1\,{}^{3}P_{2}$ & 5.840(0.6) & 5.805 & 9.0 \\ 
$B_{c}^{-}:b\overline{c}\ 1\,{}^{1}S_{0}$ & 6.276(21.) & 6.251 & 0.4 \\ 
\hline
\end{tabular}%
\end{equation*}%
display very good results for the ground states. \ Finally for the $b\bar{b}$
mesons, even though the overall fit is a good (one exception is the 3rd
radial excitation), the spin -- spin splitting of the ground state is,
oddly, not as good as for the lighter mesons.\bigskip\ 
\begin{equation*}
\begin{tabular}{|lccc|}
\hline
$b\bar{b}$ Mesons~~~~~~~~~~~~ & Exp. & Th. & $\chi ^{2}$-Th. \\ \hline
$\eta _{b}:b\overline{b}\ 1\,{}^{1}S_{0}$ & 9.389(4.0) & 9.330 & 2.0 \\ 
$\Upsilon (1S):b\overline{b}\ 1\,{}^{3}S_{1}$ & 9.460(0.3) & 9.444 & 2.6 \\ 
$\chi _{b0}:b\overline{b}\ 1\,{}^{3}P_{0}$ & 9.859(0.4) & 9.834 & 5.6 \\ 
$\chi ~_{b1}:b\overline{b}\ 1\,{}^{3}P_{1}$ & 9.893(0.3) & 9.886 & 0.4 \\ 
$\chi _{b2}:b\overline{b}\ 1\,{}^{3}P_{2}$ & 9.912(0.3) & 9.920 & 0.6 \\ 
$\Upsilon (2S):b\overline{b}\ 2\,{}^{3}S_{1}$ & 10.023(0.3) & 10.022 & 0.0
\\ 
$\Upsilon (D):b\overline{b}\ 2\,{}^{3}D_{2}$ & 10.161(0.6) & 10.179 & 2.3 \\ 
$\chi _{b0}:b\overline{b}\ 2\,{}^{3}P_{0}$ & 10.232(0.4) & 10.229 & 0.1 \\ 
$\chi _{b1}:b\overline{b}\ 2\,{}^{3}P_{1}$ & 10.255(0.5) & 10.262 & 0.4 \\ 
$\chi _{b2}:b\overline{b}\ 2\,{}^{3}P_{2}$ & 10.269(0.4) & 10.286 & 2.5 \\ 
$\Upsilon (3S):b\overline{b}\ 3\,{}^{3}S_{1}$ & 10.355(0.6) & 10.368 & 1.2
\\ 
$\Upsilon (4S):b\overline{b}\ 4\,{}^{3}S_{1}$ & 10.579(1.2) & 10.633 & 11.7
\\ 
$\Upsilon (5S):b\overline{b}\ 5\,{}^{3}S_{1}$ & 10.865(8.0) & 10.857 & 0.0
\\ 
$\Upsilon (6S):b\overline{b}\ 6\,{}^{3}S_{1}$ & 11.019(8.0) & 11.055 & 0.2
\\ \hline
\end{tabular}%
\end{equation*}

\bigskip

\subsection{Application of Two Body Dirac Equations to Baryon Spectroscopy}

Sazdjian \cite{saz89} combined three pairs of interacting quarks into a
single relativistically covariant three body equation for bound states,
having a Schr\"{o}dinger-like structure. There is no space to develope his
approach here. \ We say a few words about the analogy of his results to that
of the two body equations. Recall that for two bodies we have the results%
\begin{eqnarray*}
\mathcal{H}_{1}\psi &=&\left[ p_{1}^{2}+m_{1}^{2}+\Phi _{12}\right] \psi =0,
\\
\mathcal{H}_{2}\psi &=&\left[ p_{2}^{2}+m_{2}^{2}+\Phi _{12}\right] \psi =0,
\\
\varepsilon _{1} &=&[w+\left( m_{1}^{2}-m_{2}^{2}\right) /(\varepsilon
_{1}+\varepsilon _{2})]/2, \\
~~\varepsilon _{2} &=&[w+\left( m_{2}^{2}-m_{1}^{2}\right) /(\varepsilon
_{1}+\varepsilon _{2})]/2,~~~~ \\
\varepsilon _{1}+\varepsilon _{2} &=&w, \\
\left[ \mathcal{H}_{1},\mathcal{H}_{2}\right] \psi &=&0\rightarrow \Phi
_{12}=\Phi _{12}(x_{12\bot }), \\
\left( p_{\bot }^{2}+\Phi _{12}\right) \psi &=&(\varepsilon
_{1}^{2}-m_{1}^{2})\psi =(\varepsilon _{2}^{2}-m_{2}^{2})\psi =b^{2}(w)\psi .
\end{eqnarray*}

\bigskip For three bodies, speaking heuristically%
\begin{eqnarray*}
\mathcal{H}_{1}\psi &=&\left[ p_{1}^{2}+m_{1}^{2}+\Phi _{12}+\Phi _{31}%
\right] \psi =0,~ \\
\mathcal{H}_{2}\psi &=&\left[ p_{2}^{2}+m_{2}^{2}+\Phi _{23}+\Phi _{12}%
\right] \psi =0, \\
\mathcal{H}_{3}\psi &=&\left[ p_{3}^{2}+m_{3}^{2}+\Phi _{31}+\Phi _{23}%
\right] \psi =0,~ \\
\varepsilon _{1} &=&[w+\left( m_{1}^{2}-m_{2}^{2}\right) /(\varepsilon
_{1}+\varepsilon _{2})+\left( m_{1}^{2}-m_{3}^{2}\right) /(\varepsilon
_{1}+\varepsilon _{3})]/3, \\
\varepsilon _{2} &=&[w+\left( m_{2}^{2}-m_{3}^{2}\right) /(\varepsilon
_{2}+\varepsilon _{3})+\left( m_{2}^{2}-m_{1}^{2}\right) /(\varepsilon
_{2}+\varepsilon _{1})]/3, \\
\varepsilon _{3} &=&[w+\left( m_{3}^{2}-m_{1}^{2}\right) /(\varepsilon
_{3}+\varepsilon _{1})+\left( m_{3}^{2}-m_{1}^{2}\right) /(\varepsilon
_{3}+\varepsilon _{1})]/3, \\
\varepsilon _{1}+\varepsilon _{2}+\varepsilon _{3} &=&w \\
\Phi _{12} &=&\Phi _{12}(x_{12\bot }),\Phi _{23}=\Phi _{23}(x_{23\bot
}),\Phi _{31}=\Phi _{31}(x_{31\bot }), \\
x_{ij\perp }^{\mu } &=&(x_{i}^{\mu }-x_{j}^{\mu })+\hat{P}^{\mu }\hat{P}%
\cdot (x_{i}-x_{j}),
\end{eqnarray*}%
where $P=\sum_{i=1}^{N}p_{i}$,is the \textit{total }momentum (not $%
p_{i}+p_{j}$). \ Unlike the case of two bodies, the $x_{ij\perp }^{\mu }$
dependence is obained by a more roundabout approach. \ 

The sum three body Schr\"{o}dinger-like which we adopt from his approach is%
\cite{whit} \ 
\begin{eqnarray}
\mathcal{H\psi } &\equiv &\frac{1}{F}\left( \frac{p_{1\perp }^{2}+\Phi
_{12}+\Phi _{13}}{2\varepsilon _{1}(w,m_{1},m_{2},m_{3})}+\frac{p_{2\perp
}^{2}+\Phi _{23}+\Phi _{12}}{2\varepsilon _{2}(w,m_{1},m_{2},m_{3})}+\frac{%
p_{3\perp }^{2}+\Phi _{31}+\Phi _{23}}{2\varepsilon _{3}(w,m_{1},m_{2},m_{3})%
}\right) \psi  \notag \\
&=&(w-m_{1}-m_{2})\psi ,
\end{eqnarray}%
in which $F=F(w,m_{1},m_{2},m_{3})$ is a complicated function \ of the
invariant $w$ and the three masses. \ We choose $\Phi _{ab}$ to have the
same functional dependence on $S$ and $A$ as in SLE form of TBDE 
\begin{align*}
& \!\!\!\!\!\!\!\!\!\!\!\!\!\!\Phi _{ab}(\mathbf{r}_{ab}\mathbf{,}%
m_{a},m_{b},w_{ab},\mathbf{\sigma }_{a},\mathbf{\sigma }_{b}) \\
=& 2m_{w_{ab}}S+S^{2}+2\varepsilon _{w_{ab}}A-A^{2}+2\varepsilon
_{w_{ab}}V-V^{2}+\Phi _{D} \\
& +\mathbf{L}_{ab}\mathbf{\cdot (\sigma }_{a}\mathbf{+\sigma }_{b}\mathbf{)}%
\Phi _{SO}+\mathbf{\sigma }_{a}\mathbf{\cdot \hat{r}}_{ab}\mathbf{\sigma }%
_{b}\mathbf{\cdot \hat{r}}_{ab}\mathbf{L}_{ab}\mathbf{\cdot (\sigma }_{a}%
\mathbf{+\sigma }_{b}\mathbf{)}\Phi _{SOT} \\
& +\mathbf{\sigma }_{a}\mathbf{\cdot \sigma }_{b}\Phi _{SS}+(3\mathbf{\sigma 
}_{a}\mathbf{\cdot \hat{r}}_{ab}\mathbf{\sigma }_{b}\mathbf{\ \cdot \hat{r}}%
_{ab}\mathbf{-\sigma }_{a}\mathbf{\cdot \sigma }_{b})\Phi _{T} \\
& +\mathbf{L}_{ab}\mathbf{\cdot (\sigma }_{a}\mathbf{-\sigma }_{b}\mathbf{)}%
\Phi _{SOD}+i\mathbf{L}_{ab}\mathbf{\cdot \sigma }_{a}\mathbf{\times \sigma }%
_{b}\Phi _{SOX}, \\
w_{ab}=& \varepsilon _{a}+\varepsilon _{b}.
\end{align*}

\ Note in the NR limit, $F\rightarrow 1$ and $\varepsilon _{i}\rightarrow
m_{i}$. This spin-dependent potential used in the three-body bound state
equation is not a result of the reduction of some set of three-body Dirac
equations. Rather, it is the two-body SLE quasipotential inserted by hand as
an addition into free Klein-Gordon forms. The equation is solved by
variational approach\cite{whit}. \ \ 

For the ground state octet the spectral results below (one set of parameters
for the entire baryon spectrum) indicate a good fit for the nucleons, high
for the strangeness caring baryons but low for the $\Lambda .$

\begin{equation*}
\begin{tabular}{|c|c|c|c|c|c|c|}
\hline
Baryon & $J$ & $L$ & $S$ & Th. Mass (MeV) & Exp. Mass(MeV) & Exp-Th.(MeV) \\ 
\hline
$p$ & 1/2 & 0 & 1/2 & 947 & 938 & -9 \\ \hline
$n$ & 1/2 & 0 & 1/2 & 948 & 939 & -9 \\ \hline
$\Sigma ^{+}$ & 1/2 & 0 & 1/2 & 1250 & 1189 & -61 \\ \hline
$\Sigma ^{0}$ & 1/2 & 0 & 1/2 & 1261 & 1192 & -68 \\ \hline
$\Sigma ^{-}$ & 1/2 & 0 & 1/2 & 1271 & 1197 & -73 \\ \hline
$\Xi ^{0}$ & 1/2 & 0 & 1/2 & 1373 & 1314 & -58 \\ \hline
$\Xi ^{-}$ & 1/2 & 0 & 1/2 & 1378 & 1321 & -57 \\ \hline
$\Lambda ^{0}$ & 1/2 & 0 & 1/2 & 1082 & 1125 & 43 \\ \hline
\end{tabular}%
\end{equation*}

For the ground state decimet the higher strangeness particles lie lower
instead of higher as with the octet

\begin{equation*}
\begin{tabular}{|c|c|c|c|c|c|c|}
\hline
Baryon & $J$ & $L$ & $S$ & Th. Mass (MeV) & Exp. Mass(MeV) & Exp-Th.(MeV) \\ 
\hline
$\Delta ^{++}$ & 3/2 & 0 & 3/2 & 1249 & 1232 & -17 \\ \hline
$\Delta ^{+}$ & 3/2 & 0 & 3/2 & 1250 & 1232 & -18 \\ \hline
$\Delta ^{0}$ & 3/2 & 0 & 3/2 & 1251 & 1232 & -19 \\ \hline
$\Delta ^{-}$ & 3/2 & 0 & 3/2 & 1252 & 1232 & -20 \\ \hline
$\Sigma ^{+}(1390)$ & 3/2 & 0 & 3/2 & 1384 & 1383 & -1 \\ \hline
$\Sigma ^{0}(1390)$ & 3/2 & 0 & 3/2 & 1385 & 1384 & -1 \\ \hline
$\Sigma ^{-}(1390)$ & 3/2 & 0 & 3/2 & 1387 & 1387 & 0 \\ \hline
$\Xi ^{0}(1530)$ & 3/2 & 0 & 3/2 & 1501 & 1531 & 30 \\ \hline
$\Xi ^{-}(1530)$ & 3/2 & 0 & 3/2 & 1507 & 1535 & 28 \\ \hline
$\Omega ^{-}$ & 3/2 & 0 & 3/2 & 1609 & 1672 & 63 \\ \hline
\end{tabular}%
\end{equation*}

\bigskip For the orbital and radial excitations the results are mixed but
special note is taken for the good fit to the $\Lambda (1405)$

\begin{equation*}
\begin{tabular}{|c|c|c|c|c|c|c|}
\hline
Baryon & $J$ & $L$ & $S$ & Th. Mass (MeV) & Exp. Mass(MeV) & Exp-Th. \\ 
\hline
$N(1440)$ & 1/2 & 0 & 1/2 & 1557 & 1420-1470 & -117 \\ \hline
$\Lambda (1600)$ & 1/2 & 0 & 1/2 & 1677 & 1560-1700 & -77 \\ \hline
$\Sigma (1660)$ & 1/2 & 0 & 1/2 & 1672 & 1630-1690 & 12 \\ \hline
$\Sigma (1880)$ & 1/2 & 0 & 1/2 & 1709 & 1800-1960 & 171 \\ \hline
$\Xi (1690)$ & 1/2 & 0 & 1/2 & 1784 & 1680-1700 & -94 \\ \hline
$\Delta (1600)$ & 3/2 & 0 & 3/2 & 1521 & 1550-1700 & 78 \\ \hline
$N(1535)$ & 1/2 & 1 & 1/2 & 1549 & 1525-1545 & -14 \\ \hline
$\Delta (1620)$ & 1/2 & 1 & 1/2 & 1542 & 1600-1660 & 78 \\ \hline
$\Lambda (1405)$ & 1/2 & 1 & 1/2 & 1410 & 1402-1410 & -4 \\ \hline
$\Lambda (1670)$ & 1/2 & 1 & 1/2 & 1671 & 1660-1680 & -1 \\ \hline
Baryon & $J$ & $L$ & $S$ & Th. Mass (MeV) & Exp. Mass(MeV) & Exp-Th. \\ 
\hline
$N(1650)$ & 1/2 & 1 & 3/2 & 1566 & 1645-1670 & 84 \\ \hline
$\Sigma (1750)$ & 1/2 & 1 & 3/2 & 1644 & 1730-1800 & 121 \\ \hline
$\Lambda (1800)$ & 1/2 & 1 & 3/2 & 1658 & 1720-1850 & 142 \\ \hline
$N(1520)$ & 3/2 & 1 & 1/2 & 1551 & 1515-1525 & -31 \\ \hline
$\Delta (1700)$ & 3/2 & 1 & 1/2 & 1546 & 1670-1750 & 154 \\ \hline
$\Sigma (1670)$ & 3/2 & 1 & 1/2 & 1679 & 1665-1685 & -4 \\ \hline
$\Lambda (1520)$ & 3/2 & 1 & 1/2 & 1680 & 1518-1521 & -160 \\ \hline
$\Lambda (1690)$ & 3/2 & 1 & 1/2 & 1670 & 1685-1695 & 20 \\ \hline
$\Xi (1820)$ & 3/2 & 1 & 1/2 & 1777 & 1818-1828 & 43 \\ \hline
$N(1700)$ & 3/2 & 1 & 3/2 & 1568 & 1650-1750 & 132 \\ \hline
$\Sigma (1775)$ & 5/2 & 1 & 3/2 & 1661 & 1770-1780 & 114 \\ \hline
N(1675) & 5/2 & 1 & 3/2 & 1615 & 1670-1680 & 59 \\ \hline
$\Lambda (1830)$ & 5/2 & 1 & 3/2 & 1641 & 1810-1830 & 189 \\ \hline
$\Xi (1950)$ & 5/2 & 1 & 3/2 & 1757 & 1935-1965 & 192 \\ \hline
\end{tabular}%
\end{equation*}

\ Finally we have the baryons that involve the charmed and bottom quarks
with mixed results.

\begin{equation*}
\begin{tabular}{|c|c|c|c|c|c|c|}
\hline
Baryon & $J$ & $L$ & $S$ & Th. Mass (MeV) & Exp. Mass(MeV) & Exp-Th.(MeV) \\ 
\hline
$\Sigma _{c}^{++}(2455)$ & 1/2 & 0 & 1/2 & 2385 & 2454 & 68 \\ \hline
$\Sigma _{c}^{++}(2520)$ & 3/2 & 0 & 3/2 & 2551 & 2520 & -31 \\ \hline
$\Lambda _{c}^{+}(2286)$ & 1/2 & 0 & 1/2 & 2382 & 2286 & -96 \\ \hline
$\Lambda _{c}^{+}(2595)$ & 1/2 & 1 & 1/2 & 2415 & 2595 & 180 \\ \hline
$\Xi _{c}^{+}(2467)$ & 1/2 & 0 & 1/2 & 2561 & 2467 & -94 \\ \hline
$\Xi _{c}^{0}(2470)$ & 1/2 & 0 & 1/2 & 2562 & 2470 & -92 \\ \hline
$\Xi _{c}^{+}(2645)$ & 3/2 & 0 & 3/2 & 2598 & 2645 & 46 \\ \hline
$\Xi _{c}^{+}(2790)$ & 1/2 & 1 & 3/2 & 2661 & 2790 & 129 \\ \hline
$\Xi _{c}^{+}(2815)$ & 3/2 & 1 & 3/2 & 2707 & 2815 & 108 \\ \hline
$\Omega _{c}^{0}(2695)$ & 1/2 & 0 & 1/2 & 2732 & 2695 & -37 \\ \hline
$\Omega _{c}^{0}(2770)$ & 3/2 & 0 & 3/2 & 2745 & 2770 & 25 \\ \hline
$\Sigma _{b}^{+}(5829)$ & 3/2 & 0 & 3/2 & 5800 & 5829 & 29 \\ \hline
$\Sigma _{b}^{-}(5836)$ & 3/2 & 0 & 3/2 & 5851 & 5836 & -15 \\ \hline
$\Xi _{b}^{0}(5790)$ & 1/2 & 0 & 1/2 & 5854 & 5790 & -64 \\ \hline
$\Omega _{b}^{-}(6071)$ & 1/2 & 0 & 1/2 & 6032 & 6071 & 39 \\ \hline
\end{tabular}%
\end{equation*}

\subsection{Two Body Dirac Equations for QED}

The SLE given in Eq. (\ref{SLE}) can be used for QED as well as QCD bound
states. \ For meson spectroscopy, the three invariant functions $S(r),~A(r),$
and $V(r)$ fix $\Phi (\mathbf{r,}m_{1},m_{2},w,\mathbf{\sigma }_{1},\mathbf{%
\sigma }_{2})$. \ For QED bound states $S(r)=V(r)=0$ and 
\begin{equation*}
A(r)=-\frac{\alpha }{r}\equiv -\frac{\alpha }{\sqrt{x_{\perp }^{2}}}=-\frac{%
\alpha }{\left\vert \mathbf{r}\right\vert }(\text{in c.m.).}
\end{equation*}%
The QED spectral results follow from solving numerically or analytically,
the radial forms of SLE. \ For equal mass spin singlet, the attractive
spin-spin quasipotential ($-3\Phi _{SS}$) exactly cancels repulsive Darwin
quasipotential $\Phi _{D}$, giving an eigenvalue equation for $^{1}J_{J}$
states

\begin{equation*}
\{-\frac{d^{2}}{dr^{2}}+\frac{J(J+1)}{r^{2}}+2\varepsilon
_{w}A-A^{2}\}u_{0}=b^{2}u_{0}.
\end{equation*}%
For point electron and positron $A=-\alpha /r$ $\rightarrow $

\begin{equation*}
\{-\frac{d^{2}}{dr^{2}}+\frac{J(J+1)}{r^{2}}-\frac{2\varepsilon _{w}\alpha }{%
r}-\frac{\alpha ^{2}}{r^{2}}\}u_{0}=b^{2}u_{0}.
\end{equation*}%
This has ground state analytic spectral solution \cite{exct} with accepted $%
O(\alpha ^{4})$ perturbative expansion%
\begin{equation*}
w=m\sqrt{2+2/\sqrt{1+\frac{\alpha ^{2}}{\left( \frac{1}{2}+\sqrt{\frac{1}{4}%
-\alpha ^{2}}\right) ^{2}}}}=2m-\frac{m\alpha ^{2}}{4}-\frac{21m\alpha ^{4}}{%
64}+O(\alpha ^{6})..,
\end{equation*}%
At short distance, the SLE equation takes on the limiting form 
\begin{equation*}
\{-\frac{d^{2}}{dr^{2}}+\frac{J(J+1)}{r^{2}}-\frac{\alpha ^{2}}{r^{2}}\}u=0.
\end{equation*}%
$\ \ $Since $J(J+1)-\alpha ^{2}$ $>-1/4~$\ the effective potential is
nonsingular, implying a well defined solutions. \ Numerical solutions of the
eigenvalue eqations yield spectra agreeing with standard perturbative $%
O(\alpha ^{4})$ results. \ E.g. for the singlet ground state of positronium%
\cite{becker} 
\begin{eqnarray*}
\text{numerical binding energy} &=&-6.8033256279~\text{eV}, \\
\text{vs~}m(-\alpha ^{2}/4-21\alpha ^{4}/64) &=&-6.8033256719~\text{eV.}
\end{eqnarray*}%
The difference is on order of $m\alpha ^{6}.~$For the triplet ground state
of positronium 
\begin{eqnarray*}
\text{numerical binding energy} &=&-6.8028426132~\text{eV}, \\
\text{vs }m(-\alpha ^{2}/4+\alpha ^{4}/192) &=&-6.8028426636~\text{eV.}
\end{eqnarray*}%
The difference is also on order of $m\alpha ^{6}$. This does not include
annihilation diagram (nor radiative corrections). \ 

These two results are from a very extensive list of numerically computed
spectral \cite{becker} showing TBDE passes crucial tests, ones not
demonstrated in any other relativistic bound state formalism. Sommerer,
Spence and Vary \cite{vary} have found a particular quasipotential formalism
that does give such agreement, but only for the ground state. They also
demonstrate that several prominent two-body relativistic bound state
formalisms (including the Blankenbecler-Sugar formalism \cite{sugar}, and
the formalism of Gross \cite{gross})\ fail this important test. The
importance of numerical tests of the formalisms is not for QED, but rather
as a reliability test for use of the corresponding formalisms, e.g. Coulomb
gauge BSE in QCD \cite{swan}. \ If failure occurs in their applications to
QED bound states this brings into question the spectral results of similar
nonperturbative (i.e. numerical) approaches applied to QCD bound states.

\subsubsection{Peculiar Singlet Positronium Bound States}

These last two topics are on peculiar solutions of the TBDE and are
speculative with new phenomena predicted for the positronium system. \ We
begin by a critical examination of the bound state equation for point $e^{+}$
and $e^{-}.$%
\begin{equation}
\{-\frac{d^{2}}{dr^{2}}-\frac{2\varepsilon _{w}\alpha }{r}-\frac{\alpha ^{2}%
}{r^{2}}\}u_{0}=b^{2}u_{0}  \label{exct}
\end{equation}%
\ Based on this equation \cite{pec} we find: a new \ positronium bound state
with a large $\left( 300~\text{KeV}\right) $binding energy derived from an
exact solution of the above equation. The new \ positronium bound state
would result from a metastable two-photon decay of the usual positronium
ground state which has a binding energy of about $6.8$ eV. \ It then
annihilates promptly into 2 photon with c.m. energy of $700~$KeV. \ The
existence of this new positronium state would thus be a distinctive 4 photon
decay signature of the usual singlet positronium ground state. \ The size of
the new positronium bound state is on the order of an electron's Compton
wave length. \ 

\bigskip Eq .(\ref{exct}) has the short distance ($r<<\alpha /2\varepsilon
_{w})~$behavior 
\begin{equation*}
\left\{ -\frac{d^{2}}{dr^{2}}-\frac{\alpha ^{2}}{r^{2}}\right\} u=0,
\end{equation*}%
with solutions called usual and peculiar, 
\begin{eqnarray*}
u_{+} &\sim &r^{\lambda _{+}+1};\text{ }\lambda _{+}=(-1+\sqrt{1-4\alpha ^{2}%
})/2;+~\text{usual} \\
u_{-} &\sim &r^{\lambda _{-}+1};\text{ }\lambda _{+}=(-1-\sqrt{1-4\alpha ^{2}%
})/2~;-~\text{peculiar.}
\end{eqnarray*}%
With these behaviors, the probability is finite for both signs 
\begin{equation*}
\psi _{\pm }^{2}d^{3}r=\frac{u_{\pm }^{2}}{r^{2}}r^{2}drd\Omega =u_{\pm
}^{2}drd\Omega =r^{(1\pm \sqrt{1-4a^{2}})}drd\Omega .
\end{equation*}
Both of these behaviors are quantum mechanically acceptable near the origin
. If $L\neq 0$ so that $L(L+1)-\alpha ^{2}>0$ or the electron is not a point
particle then the peculiar solution not physically admissible.

Both $^{1}S_{0}$ \ bound state solutions can be obtained analytically. The
respective sets of eigenvalues for total invariant c.m. energy (mass) $%
w_{\pm n}~$($n$ is principle quantum \#) \ 
\begin{equation}
w_{\pm n}=m\sqrt{2+2/\sqrt{1+{\alpha ^{2}}/{(}n\pm \sqrt{1/4-\alpha ^{2}}-1/2%
{)^{2}}}}.  \notag
\end{equation}%
The usual state ground eigenvalue gives standard QED perturbative results
thru order $\alpha ^{4}$ 
\begin{equation*}
w_{+n}=2m-m{\alpha ^{2}}/{4}-21m\alpha ^{4}/64+O(\alpha ^{6}),~n=1,2,3,...
\end{equation*}%
The peculiar ground state $n=1$ has mass 
\begin{equation*}
w_{-1}=m\sqrt{2+2/\sqrt{1+{\alpha ^{2}}/({1/2}-\sqrt{1/4-\alpha ^{2}}{)^{2}}}%
}\sim \sqrt{2}m\sqrt{1+\alpha },
\end{equation*}%
which represents very tight binding energy on order 300 KeV for an $%
e^{+}e^{-}$ state. \ Its weak coupling limit is antiintuitive, having a
total c.m. energy $\rightarrow $ $\sqrt{2}m$ instead of $2m$. \ \ \ \ \ \ \
\ \ \ \ \ \ \ 

The two $n=1$ wave functions have the respective forms 
\begin{eqnarray*}
u_{+}(r) &=&c_{+}r^{\lambda _{+}+1}\exp (-\kappa _{+}\varepsilon
_{w_{+}}\alpha r),~\kappa _{+}=\frac{2}{1+\sqrt{1-4\alpha ^{2}}}=\frac{1}{%
\lambda _{+}+1}, \\
u_{-}(r) &=&c_{-}r^{\lambda _{-}+1}\exp (-\kappa _{-}\varepsilon
_{w_{-}}\alpha r),~\kappa _{-}=\frac{2}{1-\sqrt{1-4\alpha ^{2}}}=\frac{1}{%
\lambda _{-}+1}.
\end{eqnarray*}%
Since they are both zero node solutions, they are not orthogonal (although
the inner product is small, $\sim 1/1000$)%
\begin{equation*}
\langle u_{-}|u_{+}\rangle =\int_{0}^{\infty }dru_{+}(r)u_{-}(r)\sim ~\alpha
^{3/2}\neq 0.
\end{equation*}

How do we reconcile this with the expected orthogonality of the
eigenfunctions of a self-adjoint operator corresponding to different
eigenvalues? One can show that the second derivative is not self-adjoint in
this context! \ However, we emphasize the fact that both the set of usual
and peculiar states are quantum mechanically admissible states. \ We admit
both types of physical states into a larger Hilbert space by introducing a
new observable $\hat{\zeta}$ with a quantum number which we call
"peculiarity" allowing the mass operator to be self-adjoint, and the set of
physically allowed states become a complete set. In particular such that

\begin{eqnarray*}
\hat{\zeta}\chi _{+} &=&\zeta \chi _{+}~~\mathrm{with~eigenvalue~}\zeta =+1,%
\text{ usual positronium,} \\
\hat{\zeta}\chi _{-} &=&\zeta \chi _{-}~~\mathrm{with~eigenvalue~}\zeta =-1,%
\text{ peculiar positronium,}
\end{eqnarray*}%
with the corresponding spinor wave function $\chi _{\zeta }$ assigned to the
states so that a usual state is represented by the peculiarity spinor $\chi
_{+}$, 
\begin{equation*}
\chi _{+}=%
\begin{pmatrix}
1 \\ 
0%
\end{pmatrix}%
,
\end{equation*}%
and a peculiar state is represented by the peculiarity spinor $\chi _{-}$ 
\begin{equation*}
\chi _{-}=%
\begin{pmatrix}
0 \\ 
1%
\end{pmatrix}%
.
\end{equation*}

With this introduction, a general wave function can be expanded in terms of
the complete set of basis functions $\{u_{+n},u_{-n}\}$ as%
\begin{equation*}
\Psi =\sum_{\zeta n}a_{\zeta n}u_{\zeta n}\chi _{\zeta },
\end{equation*}
where $n$ represents spin and spatial quantum numbers and $\zeta $ the
peculiarity. \ The variational principle applied to 
\begin{equation*}
\langle H\rangle =\frac{\langle \Psi |H|\Psi \rangle }{\langle \Psi |\Psi
\rangle },
\end{equation*}%
would lead to 
\begin{eqnarray*}
Hu_{+n}\chi _{+} &=&-\kappa _{+n}^{2}u_{+n}\chi _{+}, \\
Hu_{-n}\chi _{-} &=&-\kappa _{-n}^{2}u_{-n}\chi _{-}.
\end{eqnarray*}%
Thus the introduction of the peculiarity quantum number resolves the problem
of the over-completeness property of the basis states and the
non-self-adjoint property of the mass operator.

If the peculiarity quantum number is strictly conserved it would be
impossible for the usual positronium ground state ($1S_{u}$) to decay to the
peculiar ground state ($1S_{p}$)$\ $and usual ground state would only
undergo the usual two photon annihilation in about $10^{-10}$sec. We
consider possible evidence that this quantum number is not conserved for the
full Hamiltonian. In that case we could have that the usual ground state
undergo a metastable decay into the peculiar ground state by emitting two
photons. We obtain a lifetime of%
\begin{equation}
\tau _{1S_{u}~\rightarrow ~1S_{p}+2\gamma }\sim \frac{\tau
_{1S_{u}~\rightarrow ~2\gamma ~}\pi ^{4}}{2.55\alpha ^{2}}=9.0\times 10^{-5}%
\text{sec.}
\end{equation}
\ and a two photon annihilation lifetime $1S_{p}$ on the order of 
\begin{equation*}
\tau _{1S_{p}\rightarrow 2\gamma }\sim \frac{\tau _{1S_{u}\rightarrow
2\gamma }}{\alpha ^{3}}\symbol{126}10^{-16}\text{sec.}
\end{equation*}

This implies that we would see $4\gamma $ as the signature of the production
and decay of the peculiar positronium ground state. We obtain a small
branching ratio compared \ with the annihilation of the usual positronium
ground state into two 500 KeV photons. Failure to find the peculiar state at
the predicted energy would imply that electron and positron are not
point-like or that radiative corrections lead to less attractive potentials
that do not give quantum mechanically acceptable double roots of the leading
short distance behavior.

\subsubsection{New Peculiar $^{3}P_{0}$ $e^{+}e^{-\text{ }}$ QED Resonances}

A closely related state to peculiar positronium is a pure QED $e^{+}~e^{-}$
resonance from the highly attractive magnetic spin-orbit interaction between
point electron\ and positron in the $^{3}P_{0}~$angular momentum state. \ We
find in \ particular a resonance at about 28 MeV with a narrow width of \
abouit 15 KeV\cite{pec}.

The angular momentum barrier is overwhelmed by relativistic effective
potentials at very short distances.\ \ The \ SLE for the $^{3}P_{0}$ state is

\begin{eqnarray}
\left\{ -\frac{d^{2}}{dr^{2}}+\frac{2}{r^{2}}+\Phi (r)\right\} u &=&b^{2}u, 
\notag \\
\frac{2}{r^{2}}+\Phi (r) &=&\frac{2}{(r+2\alpha /w)^{2}}-\frac{2\varepsilon
_{w}\alpha }{r}-\frac{\alpha ^{2}}{r^{2}}.  \label{5}
\end{eqnarray}%
In first term on the right hand side, we see that the angular momentum
barrier $2/r^{2}$ is overwhelmed by the net effects of the magnetic
interactions (spin-orbit, spin-spin, tensor and Darwin interactions) and the 
$-\alpha ^{2}/r^{2}$ portion embodied in $\Phi $ at a radius of about $%
2\times 10^{-3}$ fermis. At short distances, the effective potential is
highly attractive ($\sim -\alpha ^{2}/r^{2}$) but not technically singular.
\ Before going on to the solution of this equation for scattering states, we
examine scattering solutions of the $^{1}S_{0}$ state.

\ The radial SLE is%
\begin{equation*}
\{-\frac{d^{2}}{dr^{2}}-\frac{2\varepsilon _{w}\alpha }{r}-\frac{\alpha ^{2}%
}{r^{2}}\}u=b^{2}(w)u=\frac{1}{4}(w^{2}-4m^{2})u.
\end{equation*}%
~For For scattering states, the Coulomb term and $-\alpha ^{2}/r^{2}$ lead
to the exact relativistic Coulomb wave functions%
\begin{align*}
\bar{u}& =aF_{\lambda }(\eta ,br)+cG_{\lambda }(\eta ,br), \\
\lambda (\lambda +1)& =-\alpha ^{2},\text{ }\lambda _{\pm }=\frac{1}{2}%
(-1\pm \sqrt{1-4\alpha ^{2}}) \\
\eta & =-\frac{\varepsilon _{w}\alpha }{b}.
\end{align*}%
The lower sign correspond to peculiar solutions and the upper to the usual
solutions. \ \ The asymptotic behavior of the regular Coulomb wave function
is

\ \ 
\begin{equation*}
F_{\lambda _{\pm }}(\eta ,br\rightarrow \infty )\rightarrow \mathrm{const}%
\times \sin (br-\eta \log 2br+\sigma _{\lambda _{\pm }}-\lambda _{\pm }\pi
/2).
\end{equation*}

~Two roots gives two sets of Coulomb phase shifts.

\begin{eqnarray*}
\delta _{\lambda _{\pm }} &=&\sigma _{\lambda _{\pm }}-\lambda _{\pm }\pi
/2,~ \\
~\sigma _{\lambda _{\pm }} &=&\eta \psi (\lambda _{\pm
}+1)+\sum_{n=0}^{\infty }\left( \frac{\eta }{\lambda _{\pm }+1+n}-\arctan (%
\frac{\eta }{\lambda _{\pm }+1+n})\right)
\end{eqnarray*}

\ \ How might Eq. (\ref{5}) lead to a resonance? The short distance behavior
($r<<2\alpha /w)$ has the same usual and peculiar solutions as for the $%
^{1}S_{0}$ state. We solve for the phase shift by the phase method of
Calogero giving a nonlinear equation for the phase shift function. \
Starting with boundary conditions and integrating to infinity gives the
phase shift. Built into the solutions are the Coulomb and negative barrier
terms so that the equation is for the residual phase shift function due just
to the real barrier and magnetic spin terms

\begin{eqnarray*}
\gamma _{\pm }^{\prime }(r) &=&-\frac{2}{b(r+2\alpha /w)^{2}}(\cos \gamma
_{\pm }(r)F_{\lambda _{\pm }}(r)+\sin \gamma _{\pm }(r)G_{\lambda _{\pm
}}(r))^{2}, \\
\gamma _{\pm }(0) &=&0.
\end{eqnarray*}%
From this we obtain the total phase shift

\begin{equation*}
\delta =\delta _{1}+\sigma _{1}=\gamma _{\pm }(\infty )+\sigma _{\lambda
_{\pm }}+(1-\lambda _{\pm })\pi /2.
\end{equation*}
This leads to no resonance for any energy for usual solution $\lambda _{+}= 
\frac{1}{2}(-1+\sqrt{1-4\alpha ^{2}})$ and a 28 MeV resonance of 15 keV
width for the peculiar solution \ $\lambda _{-}=\frac{1}{2}(-1-\sqrt{
1-4\alpha ^{2}}).$ \ The resonance disappears if the electron and positrons
are not point particles.

\ \ \ \ \ \ \ \ \ \ \ \ \ \ \ \ \ \ \ \ \ \ \ \ \ \ \ \ \ \ \ \ \ \ \ \ \ \
\ \ \ \ \ \ \ \ \ \ \ \ \ \ \ \ \ \ \ \ \ \ \ \ \ \ \ \ 

\section{\protect\bigskip Summary}

The \ Two Body Dirac equations of constraint dynamics have dual origins in
QFT and the classical relativistic two body problem. \ With the Adler-Piran
potential the TBDE gives a very good fit to entire meson spectrum with just
3 invariant functions $A(r),V(r),S(r)$. We use the TBDE in the three
two-body subsystems for baryon spectroscopy (we have not yet examined the
three-body Dirac equations). The nonperturbative structure of the TBDE makes
it more than competitive with other approaches since its QED applications
reproduce numerically known perturbative spectrum. \ Finally, assuming
point-like electron and positron, the TBDE predict new and peculiar
positronium bound states and resonances.

\bigskip


\bigskip

\begin{thebibliography}{99}
\bibitem{saz} H. Sazdjian, J. Math. Phys. \textbf{28} 2618 (1987); \textit{%
Extended Objects and Bound Systems}, Proceedings of the Karuizawa
International Symposium, 1992, eds. O. Hara, S. Ishida, and S. Nake (World
Scientific, Singapore, 1992), p 117.

\bibitem{cra82} P. Van Alstine and H.W. Crater, J. Math. Phys. \textbf{23},
1997 (1982); H. W. Crater and P. Van Alstine, Ann. Phys. (N.Y.) \textbf{148}
, 57 (1983).

\bibitem{jim} Horace W. Crater and James Schiermeyer, Phys.Rev .\textbf{D 82}%
:094020, 2010.

\bibitem{yin} Horace W. Crater, Jin-Hee Yoon and Cheuk-Yin Wong, Phys. Rev. 
\textbf{D 79}:034011,2009.

\bibitem{whit} J. Whitney and H. Crater, Phys. Rev. D 89, 014123 (2014) \ 

\bibitem{becker} H. W. Crater, R. L. Becker, C. Y. Wong, and P. Van Alstine,
Phys. Rev. \textbf{D46} , 5117 (1992).

\bibitem{pec} H. Crater and C. Y. Wong, Phys. Rev. D 85, 116005 (2012).

\bibitem{bse} E. E. Salpeter and H. A. Bethe, Phys. Rev. \textbf{84} 1232,
(1951),

\bibitem{wick} G. C. Wick, Phys. Rev. \textbf{96},1124 (1954) and R. E.
Cutkosky, Phys. Rev. \textbf{96},1135 (1954)

\bibitem{nak} N. Nakanishi, Suppl. Prog. Theor. Phys. \textbf{\ 43}, 1
(1969).

\bibitem{jal} H. Jallouli and H. Sazdjian, J. Phys. G \textbf{22,}1119
(1996).

\bibitem{yaes} R. Yaes, Phys. Rev.\textbf{D3},3086 (1971) .

\bibitem{tod} I. T. Todorov, Phys. Rev. \textbf{D3} , 2351 (1971).

\bibitem{saz97} H. Jollouli and H. Sazdjian, Annals of Physics \textbf{253},
376 (1997).

\bibitem{cur} D. G. Currie, T. F. Jordan, and E. C. G. Sudarshan, Rev. Mod.
Phys. \textbf{35} 350, 1032 (1963).

\bibitem{tod78} I. T. Todorov, Dubna Joint Institute for Nuclear Research
No. E2-10175, 1976; Ann. Inst. H. Poincar\'{e} \textbf{A28}, 207 (1978).

\bibitem{saz96} H. Sazdjian, J.Math.Phys. \textbf{38,} 4951 (1997).

\bibitem{adler} S. L. Adler and T. Piran, Phys. Lett., \textbf{117B}, 91
(1982) and references contained therein.

\bibitem{saz89} H. Sazdjian, Annals of Physics \textbf{191},52(1989).

\bibitem{exct} P. Van Alstine and H. W. Crater, Phys. Rev. \textbf{D~34},
1932 (1986).

\bibitem{vary} A. J. Sommerer, J. R. Spence, J. P. Vary, Phys. Rev \textbf{C
49}, 513 (1994).

\bibitem{sugar} R. Blankenbecler and R. Sugar, Phys. Rev. 142, 1051 (1966).

\bibitem{gross} F. Gross, Phys. Rev. 188, 1448 (1969).

\bibitem{swan} A. P. Szczepaniak and E. S. Swanson, Phys. Rev \textbf{D55, }%
3987, (1997).
\end{thebibliography}
\end{document}